\documentclass[10pt]{article}
\usepackage{latexsym}
\usepackage{amssymb}
\usepackage{amsmath}
\usepackage{amscd}
\usepackage{amsthm}
\usepackage[left=1.5cm,top=1.5cm,right=1.5cm,bottom=1.5cm]{geometry}
\usepackage{graphicx}
\usepackage{hyperref}
\usepackage[font={footnotesize,it}]{caption}

\begin{document}
\begin{center}
\large{\bf{Viscous Dark Energy In Bianchi Type V Space-Time}} \\
\vspace{10mm}
\normalsize{Hassan Amirhashchi}\\
\vspace{5mm}
\normalsize{Department of Physics, Mahshahr Branch, Islamic Azad University,  Mahshahr, Iran \\
E-mail:h.amirhashchi@mhriau.ac.ir} \\
\end{center}
\begin{abstract}
We study the behavior of dark energy (DE) in the scope of anisotropic Bianchi type V (BV) space-time. First, we derive Friedmann-Like Equations,
then, we compare the dark energy equation of state (EoS) parameter for viscous and non-viscous dark energy and make a correspondence between DE
and quintessence and phantom descriptions of non-viscous and viscous dark energy and reconstruct the potential of these two scalar fields.
The late time behavior of EoS parameter through a thermodynamical study has also been investigated. Finally, we investigate the conditions
under which BV space-time can be mapped into the FRW and how the bulk viscose coefficient may affect dark energy EoS parameter
of our $\omega\mbox{BV}$ Model with constraints from 28 Hubble parameter, $H(z)$, measurements at intermediate redshifts $0.07\leq z\leq 2.3$.

\end{abstract}
%\newpage
%%%%%%%%%%%%%%%%%%%%%%%%%%%%%%%%%%%%%%%%%%%%%%%%%%%%%%%%%%%%%%%%%%%%%%
%%%%%%%%%%%%%%%%%%%%%%%%%%%%%%%   SECTION 1  %%%%%%%%%%%%%%%%%%%%%%%%%
\section{Introduction}

The fact that our universe, at the present time, is experiencing an accelerating phase of it's evolution has been approved by many observations
\cite{ref1,ref2, ref3, ref4}. This seems to be enigmatic, since it shows that: (1) there must be an unknown and unusual source of
energy which behaves like anti-gravity i.e. it produces negative pressure in order to overcome the attractive force of gravity,
(2) maybe, general theory of gravity should be modified. It is worth nothing, these two scenarios could be differentiated through the 
cosmic expansion history $H(z)$ and the growth rate of cosmic large scale structure $f_{g}(z)$ \cite{ref5}.
In case of dark energy, the thing which is more interesting is the amount (density) of this component. Recent observations of type Ia
supernovae indicate that almost two-thirds of the total energy density exists in a dark energy component 
(reader is advised to see \cite{ref3,ref4,ref6} for recent review). The study of dark energy is possible either through
its equation of state parameter (EoS) $\omega^{de} = p^{de}/\rho^{de}$ (the value of EoS parameter for quintessence, 
$\Lambda$CDM and phantom scenarios is $>-1$, $=-1$, and $<-1$ respectively) or through its microphysics tha is characterized by
the sound speed ($c_{s}^2$).
Although, at the fist view, cosmological
constant $\Lambda$ seems to be appropriate candidate for the dark energy, but it encounters fine-tuning problem.
This is the reason why, different forms of dynamically changing DE with $\omega^{de} = p^{de}/\rho^{de} < -1/3$,  such as
quintessence, K-essence, tachyon, phantom, ghost condensate and quintom, etc  have been proposed in the literature.
Among these scalar fields, quintessence with EoS parameter varying as $-1<\omega^{de}<-\frac{1}{3}$ and phantom with EoS parameter
$\omega<-1$ are of more scientific interest (note that the case of $\omega^{de}\ll-1$ is ruled out by observations \cite{ref7}).
However, Since current observations show that dark energy EoS parameter could be less than $-1$ \cite{ref8,ref9}, the quintessence is ruled out
and since phantom field suffers from ultraviolet quantum instabilities \cite{ref10} can not be an appropriate DE candidate describing
region with $\omega^{de} <-1$. Nevertheless, there is another scenario in which the EoS parameter of DE could vary from quintessence to phantom
without any problem associated with scalar fields mentioned above. In this scenario which is based on Eckart theorem \cite{ref11} we consider
the DE fluid to be viscous. The possibility of a viscosity dominated late epoch of the Universe with accelerated expansion was
already mentioned by Padmanabhan and Chitre \cite{ref12}. There have been valuable works done in this regard
(for example see \cite{ref13}$-$\cite{ref17}). Recently, Velten et. al \cite{ref18} have been investigated phantom dark energy
as an effect of bulk viscosity. It is worth noting that Brevik and Gorbunova \cite{ref19} show that fluid which lie in the quintessence
region ($\omega^{de} > -1$) can reduce its thermodynamical pressure and cross the barrier $\omega^{de} = -1$, and behave
like a phantom fluid ($\omega^{de} < -1$) with the inclusion of a sufficiently large bulk viscosity.\\

FRW cosmology is based on the cosmological principle which is not exactly in consistence with the recent observations \cite{ref20, ref21} 
as these observations identify tiny variations between the intensities of the microwaves coming from different directions in the sky 
(From mathematical point of view, this means that the space-time should be anisotropic i.e. metric components are different functions of time).
On the other hand, from theoretical (philosophical) point of view the following question is reasonable: does the universe necessarily have the same
symmetries on very large scales outside the particle horizon or at early times? Therefore, to be able to compare detailed observations, we may
have to find `almost FRW` models representing a universe that is FRW-like on large scales but allowing for generic in inhomogeneities
and anisotropies arising during structure formation on a small scale. For this purpose, ``Bianchi Type Space-Times`` which
are anisotropic but homogeneous are the best. Goliath and Ellis \cite{ref22} have shown that some Bianchi models isotropise due to inflation.
For example flat and open FRW models are particular case of Bianchi type $I$, $V$ respectively.\\
As mentioned above, one can deal with the study of DE through its EoS parameter as well as its
microphysics, characterized by the sound speed ($c_{s}^2$) of perturbations to the dark energy density and pressure. In this case, as the sound speed
drops below the speed of light (i.e. $c_{s}^2<1$), dark energy inhomogeneities increase, affects both CMB and matter power spectra \cite{ref23}. 
It is worth noting that the study of a fluid model of dark energy requires considering both an equation of state parameter (EoS) and sound speed $c_{s}^2$. 
Moreover, as shown in Ref \cite{ref24} a scalar field is mathematically equivalent to a fluid with a time-dependent speed of sound. In 
this case, since at the present horizon scale the scalar field dark energy perturbations are not ignorable, dynamical dark energy is
inhomogeneous. Hence, these perturbations also affect the predicted CMB anisotropy.\\
Motivated by the situation discussed above, in this paper we consider Bianchi type V (henceforth BV) to make a detailed study of
viscous dark energy. This paper is organized as follows: the metric and the field equations are presented in Sect. $2$.
Sect. $3$ deals with the exact solutions of the field equations to obtain `almost FRW` base cosmology. The
Section $4$ deals with the study of viscous dark energy EoS parameter. In Section $5$ we make correspondence between viscous DE and scalar fields.
In section $6$, through a thermodynamical study, we investigate the late-time behavior of our DE model. In section $7$ 
we present constraints on a set of cosmological parameters of our model using 28 Hubble parameter, $H(z)$, measurements at
redshifts range $0.07\leq z\leq 2.3$, and conclude in Section $8$.
%%%%%%%%%%%%%%%%%%%%%%%%%%%%%%%%%%%%%%%%%%%%%%%%%%%%%%%%%%%%%%%%%%%%%%%%%%
%%%%%%%%%%%%%%%%%%%%%%%%%%%%%%%  SECTION 2  %%%%%%%%%%%%%%%%%%%%%%%%%%
\section{The Metric and the Field Equations}
The homogeneous and anisotropic Bianchi type-V In an orthogonal form is given by
\begin{equation}
\label{eq1} ds^{2} = - dt^{2} + A^{2} dx^{2} + e^{2\alpha x} \left[B^{2}dy^{2} + C^{2}dz^{2}\right],
\end{equation}
where the metric potentials $\rm A$, $\rm B$ and $\rm C$ are functions of cosmic time $\rm t$ alone and $\rm \alpha$ is a constant. \\

The Einstein's field equations ( in gravitational units $8\pi G = c = 1 $) read as
\begin{equation}
\label{eq2} R^{i}_{j} - \frac{1}{2} R g^{i}_{j} = T^{(m) i}_{j} +
T^{(de) i}_{j},
\end{equation}
where $T^{(m) i}_{j}$ and $T^{(de) i}_{j}$ are the energy momentum tensors of Dark matter and viscous dark energy, respectively. These are
given by
\[
  T^{m i}_{j} = \mbox{diag}[-\rho^{m}, p^{m}, p^{m}, p^{m}],
\]
\begin{equation}
\label{eq3} ~ ~ ~ ~ ~ ~ ~ ~  = \mbox{diag}[-1, \omega^{m}, \omega^{m}, \omega^{m}]\rho^{m},
\end{equation}
and
\[
 T^{de i}_{j} = \mbox{diag}[-\rho^{de}, p^{de}, p^{de}, p^{de}],
\]
\begin{equation}
\label{eq4} ~ ~ ~ ~ ~ ~ ~ ~ ~ ~ ~ ~ ~ ~ = \mbox{diag}[-1, \omega^{de}, \omega^{de},
\omega^{de}]\rho^{de},
\end{equation}
where $\rho^{m}$ and $p^{m}$ are the energy density and pressure of the perfect fluid component while $\omega^{m} = p^{m}/\rho^{m}$ is
its EoS parameter. Similarly, $\rho^{de}$ and $p^{de}$ are, respectively the energy density and pressure of the viscous DE component
while $\omega^{de}= p^{de}/\rho^{de}$ is the corresponding EoS parameter. The 4-velocity vector $u^{i} = (1, 0, 0, 0)$ is assumed to
satisfy $u^{i}u_{j} = -1$.\\

In a co-moving coordinate system ($u^{i} = \delta^{i}_{0}$), Einstein's field equations (\ref{eq2}) with (\ref{eq3})
and (\ref{eq4}) for B-V metric (\ref{eq1}) subsequently lead to the following system of equations:
\begin{equation}
\label{eq5} \frac{\ddot{B}}{B} + \frac{\ddot{C}}{C} + \frac{\dot{B}\dot{C}}{BC} - \frac{\alpha^{2}} {A^{2}}
= -\omega^{m}\rho^{m}-\omega^{de}\rho^{de},
\end{equation}
\begin{equation}
\label{eq6} \frac{\ddot{C}}{C} + \frac{\ddot{A}}{A} + \frac{\dot{C}\dot{A}}{CA} - \frac{\alpha^{2}} {A^{2}}
= -\omega^{m}\rho^{m}-\omega^{de}\rho^{de},
\end{equation}
\begin{equation}
\label{eq7} \frac{\ddot{A}}{A} + \frac{\ddot{B}}{B} + \frac{\dot{A} \dot{B}}{AB} - \frac{\alpha^{2}} {A^{2}}
= -\omega^{m}\rho^{m}-\omega^{de}\rho^{de},
\end{equation}
\begin{equation}
\label{eq8} \frac{\dot{A}\dot{B}}{AB} + \frac{\dot{A}\dot{C}}{AC} + \frac{\dot{B}\dot{C}}{BC} - \frac{3\alpha^{2}}{A^{2}}
= \rho^{m} + \rho^{de},
\end{equation}
\begin{equation}
\label{eq9} \frac{2\dot{A}}{A} - \frac{\dot{B}}{B} - \frac{\dot{C}}{C} = 0.
\end{equation}
The law of energy-conservation equation ($T^{ij}_{;j} = 0$) yields
\begin{equation}
\label{eq10} \dot{\rho}^{m} + 3(1 + \omega^{m})\rho^{m}H + \dot{\rho}^{de} +3(1 + \omega^{de})\rho^{de}H = 0.
\end{equation}
The Raychaudhuri equation is found to be
\begin{equation}
\label{eq11} \frac{\ddot{a}}{a} = \frac{1}{2}\xi\theta -
\frac{1}{6}(\rho^{m} + 3p^{m}) - \frac{1}{6}(\rho^{de} + 3p^{de})
- \frac{2}{3}\sigma^{2},
\end{equation}
where $\sigma^{2}$ is the shear scalar which is given by
\begin{equation}
\label{eq12} \sigma^{2} = \frac{1}{2}\sigma^{ij}\sigma_{ij}; ~~~
\sigma_{ij} = u_{i;j} + \frac{1}{2}(u_{i;k}u^{k}u_{j} +
u_{j;k}u^{k}u_{i}) + \frac{\theta}{3}(g_{ij} + u_{i}u_{j}),
\end{equation}
and $\theta=3H$ is the scalar expansion. Here $H$ is referred to as Hubble's parameter.

%%%%%%%%%%%%%%%%%%%%%%%%%%%%%%%%%%%%%%%%%%%%%%%%%%%%%%%%%%%%%%%%%%%%%%%%%%
%%%%%%%%%%%%%%%%%%%%%%%%%%%%%%%  SECTION 3  %%%%%%%%%%%%%%%%%%%%%%%%%%
\section{Friedmann-Like Equations}
Integrating (\ref{eq11}) and engrossing the constant of integration in $B$ or $C$, without any loss of generality, we obtain the
following relation between the metric potentials
\begin{equation}
\label{eq13} A^{2} = B C.
\end{equation}
Now, to solve Einstein's field equations (\ref{eq5}) $-$ (\ref{eq8}), we use the following technique proposed by Kumar and
Yadav \cite{ref25}. Subtracting eq. (\ref{eq5}) from eq. (\ref{eq6}), eq. (\ref{eq6}) from eq. (\ref{eq7}), and eq. (\ref{eq5}) from
eq. (\ref{eq7}) and taking second integral of each, we obtain the following three relations respectively:
\begin{equation}
\label{eq14} \frac{A}{B} = d_{1}\exp{\left(k_{1}\int{\frac{dt}{a^{3}}}\right)},
\end{equation}
\begin{equation}
\label{eq15} \frac{A}{C} = d_{2}\exp{\left(k_{2}\int{\frac{dt}{a^{3}}}\right)},
\end{equation}
and
\begin{equation}
\label{eq16} \frac{B}{C} = d_{3}\exp{\left(k_{3}\int{\frac{dt}{a^{3}}}\right)},
\end{equation}
where $d_{1}$, $d_{2}$, $d_{3}$, $k_{1}$, $k_{2}$ and $k_{3}$ are constants of integration. From (\ref{eq13})$-$(\ref{eq16}), 
the metric functions $A, B, C$ can be explicitly obtained as
\begin{equation}
\label{eq17} A(t) = a,
\end{equation}
\begin{equation}
\label{eq18} B(t) = m a\exp{\left(L \int{\frac{dt}{a^{3}}}\right)},
\end{equation}
\begin{equation}
\label{eq19} C(t) = \frac{a}{m}\exp{\left(-L \int{\frac{dt}{a^{3}}}\right)},
\end{equation}
where
\begin{equation}
\label{eq20} m = \sqrt[3]{(d_{2}d_{3})}, \; \; L = \frac{(k_{2} + k_{3})}{3}, \; \; d_{2} = d^{-1}_{1},
\; \; k_{2} = - k_{1}.
\end{equation}
Hence, we can write the general form of Bianchi type V metric as
\begin{equation}
\label{eq21} ds^{2}=-dt^{2}+a^{2}\left[dx^{2}+e^{2\alpha x}\left(m^{2}e^{2L\int{a^{-3}dt}}dy^{2}+\frac{1}{m^{2}}e^{-2L \int{a^{-3}dt}}dz^{2}\right)\right].
\end{equation}

Using eqs. (\ref{eq17})$-$(\ref{eq19}) in eqs. (\ref{eq5})$-$(\ref{eq8}), we obtain the analogue of the Friedmann
equation as
\begin{equation}
\label{eq22} 2\left(\frac{\ddot{a}}{a}\right)+\left(\frac{\dot{a}}{a}\right)^{2}+\frac{L^{2}}{a^{6}}-\frac{\alpha^{2}}{a^{2}}=-p^{m}-p^{de},
\end{equation}
\begin{equation}
\label{eq23} 3\left(\frac{\dot{a}}{a}\right)^{2}-\frac{L^{2}}{a^{6}}-\frac{3\alpha^{2}}{a^{2}}=\rho^{m}+\rho^{de},
\end{equation}
where $a=(ABC)^{\frac{1}{3}}$ is the average scale factor. One can easily re-write eqs. (\ref{eq22}) and (\ref{eq23}) in the following
compact form
\begin{equation}
\label{eq24} 2\left(\frac{\ddot{a}}{a}\right)+\frac{4L^{2}}{3a^{6}}=-\frac{1}{3}(\rho+3p),
\end{equation}
\begin{equation}
\label{eq25} \left(\frac{\dot{a}}{a}\right)^{2}-\frac{L^{2}}{3a^{6}}-\frac{\alpha^{2}}{a^{2}}= \frac{1}{3}\rho,
\end{equation}
where $p=p^{m}+p^{de}$ and $\rho=\rho^{m}+\rho^{de}$ are the total pressure and the total energy density respectively.It is worth to
mention that $\alpha$ and $L$ denote the deviation from isotropy e.g. $\alpha=L=0$ indicate flat FRW Universe. It is also interesting to note that for sufficiently large $a$, almost at the present time, the Bianchi type V space-time behaves like a flat FRW Universe.
%%%%%%%%%%%%%%%%%%%%%%%%%%%%%%%%%%%%%%%%%%%%%%%%%%%%%%%%%%%%%%%%%%%%%%%%%%%%%%%%%%%%%%%%%%%%%%%%%%%%%%%%%%%%%%%
%%%%%%%%%%%%%%%%%%%%%%%%%%%%%%%  SECTION 3  %%%%%%%%%%%%%%%%%%%%%%%%%%
\section{Dark Energy Equation of State}
To derive the general form of the equation of state (EoS) for the viscous and non-viscous dark energy (DE) in Bianchi type V space-time,
we assume that dark energy and dark matter with $\omega^{m}=0$ do not interact with each other. Therefore, we can writ the conservation
equation (\ref{eq10}) for the two dark fluids separately as
separately as
\begin{equation}
\label{eq26} \dot{\rho}^{de} + 3H(1 + \omega^{de})\rho^{de} = 0,
\end{equation}
and
\begin{equation}
\label{eq27} \dot{\rho}^{m} + 3H\rho^{m}=0.
\end{equation}
Integrating eq.(\ref{eq27})leads to
\begin{equation}
\label{eq28} \rho^{m}=\rho_{0}^{m}a^{-3}.
\end{equation}
Using eqs. (\ref{eq28})in eqs. (\ref{eq24}) and (\ref{eq25}) we obtain the energy density and pressure of non-viscous dark fluid as
\begin{equation}
\label{eq29} \rho^{de}=3H^{2}-L^{2}a^{-6}-3\alpha^{2}a^{-2}-\rho_{0}^{m}a^{-3}
\end{equation}
and
\begin{equation}
\label{eq30} p^{de}=-2\frac{\ddot{a}}{a}-H^{2}-L^{2}a^{-6}+\alpha^{2}a^{-2},
\end{equation}
respectively. Finally, we obtain the general form of the non-viscous dark energy EoS parameter (EoS) as
\begin{equation}
\label{eq31} \omega^{de}=\frac{p^{de}}{\rho^{de}}=\frac{2q-1-L^{2}a^{-6}H^{-2}+\alpha^{2}a^{-2}H^{-2}}{3-L^{2}a^{-6}H^{-2}-3\alpha^{2}a^{-2}H^{-2}-3\Omega^{m}_{0}a^{-3}},
\end{equation}
where $q=-\frac{\frac{\ddot{a}}{a}}{H^{2}}$ is the deceleration parameter and $\Omega_{0}^{m}=\frac{\rho_{0}^{m}}{3H^{2}}$ is the
current value of the DM energy density.\\

To obtain the equation of state parameter (EoS) of viscous dark energy, we assume the following expression for the pressure of the
viscous fluid \cite{ref11}
\begin{equation}
\label{eq32} p^{de}_{eff}=p^{de}+\Pi;~~~~~\omega^{de}_{eff}=p^{de}_{eff}/\rho^{de}
\end{equation}
where $\Pi = -\xi(\rho^{de})u^{i}_{;i}$ is the viscous pressure and $u^{i}_{;i}=3H $ is the covariant derivative of the
4-velocity vector $u^{i}$ . Here $\omega^{de}_{eff}$ is referred to as the effective equation of state parameter of viscous dark energy.
As noted in \cite{ref26}, since in an irreversible process the positive sign of the entropy changes, $\xi$ should be a positive parameter.
In general, $\xi(\rho^{de})=\xi_{0}(\rho^{de})^{\tau}$, where $\xi_{0}>0$ and $\tau$ are constant parameters.\\

Using eq. (\ref{eq31}) in eq. (\ref{eq32}), the EoS parameter of viscous dark energy is obtained as
\begin{equation}
\label{eq33} \omega^{de}_{eff}=\omega^{de}+\frac{\Pi}{\rho^{de}}=\frac{2q-1-L^{2}a^{-6}H^{-2}+\alpha^{2}a^{-2}H^{-2}}{3-L^{2}a^{-6}H^{-2}-3\alpha^{2}a^{-2}H^{-2}
-3\Omega^{m}_{0}a^{-3}}-3\xi_{0}\frac{H^{1-2\eta}}{(3\Omega^{de})^{\eta}},
\end{equation}
where $\Omega^{de}=\frac{\rho^{de}}{3H^{2}}$ and $\eta=1-\tau$. Note that in eq. (\ref{eq33}), $\tau$ (equivalently $\alpha$) should be
a positive number as a negative $\tau$ (or $\alpha$) force the EoS parameter to stay in phantom region for ever which is not a consistent
result (see discussion bellow).\\

Now we take a more precise look at eqs. (\ref{eq31}) and (\ref{eq33}). As we mentioned in previous section, BI-V behaves almost as flat
FRW universe at present time i. e. for very high value of the scalar factor $a$. Hence, one may characterize the universe
by $\alpha=L\sim0$ and $a\sim0$ at the current time. Therefore, we can obtain the present form of eqs. (\ref{eq31}) and (\ref{eq33})
approximately as
\begin{equation}
\label{eq34} \omega^{de}\sim\frac{2q-1}{3}
\end{equation}
and
\begin{equation}
\label{eq35} \omega^{de}_{eff}\sim\frac{2q-1}{3}-\frac{213\xi_{0}}{(12501.68)^{\alpha}}
\end{equation}
respectively.\\
According to the recent observations the deceleration parameter is restricted as $-1\leq q<0$. Applying this limit on
eqs. (\ref{eq34}) and (\ref{eq35}) we obtain
\begin{equation}
\label{eq36} -1\leq\omega^{de}<-\frac{1}{3}; \;\;\;\;\;\ -1-\frac{213\xi_{0}}{(12501.68)^{\eta}}\leq\omega^{de}_{eff}<-\frac{1}{3}-\frac{213\xi_{0}}{(12501.68)^{\eta}}.
\end{equation}
This equation clearly shows that the EoS parameter of no-viscous DE dose not cross PDL whereas the viscous dark energy equation
of state, $\omega^{de}_{eff},$ cross PDL for appropriate values of $\alpha$ and $\xi_{0}$. Consequently, non-viscous DE can only
describe the quintessence scenario of DE where $-1<\omega<-\frac{1}{3}$ and hence it is not in consistent with the recent cosmological
observations data indicating that $\omega^{de}$ is less than $-1$ today. However, since for a positive
$\tau$, $\xi(\rho^{de})=\xi_{0}(\rho^{de})^{\tau}$ is a decreasing function of time
\footnote{Note that the energy density $\rho$ is a decreasing function of time in an expanding universe.}, as time
is passing the viscosity dies out and ultimately $\omega^{de}_{eff}$ tends to the cosmological constant, $\omega^{de}=-1$, as expected.
As noted by Carroll et.al \cite{ref10}, any phantom model with $\omega^{de}<-1$ should decay to $\omega^{de}=-1$ at late time. This behavior
ensures that there is no future singularity (Big Rip); rather, the universe eventually settles into a de-Sitter phase. Therefore,
regardless the phantom model driven from the scalar fields, our phantom model generated by the bulk viscosity does not suffer
from the ultraviolet quantum instabilities as well as big rip problem.\\
%%%%%%%%%%%%%%%%%%%%%%%%%%%%%%%%%%%%%%%%%%%%%%%%%%%%%%%%%%%%%%%%%%%%%%%%%%
%%%%%%%%%%%%%%%%%%%%%%%%%%%%%%%  SECTION 2  %%%%%%%%%%%%%%%%%%%%%%%%%%
\section{Correspondence Between DE and Scalar Field}
From our above discussion we conclude that assuming the cosmic fluid to be non-viscous or viscous, one may can generate quintessence
and phantom like equation of state parameter in anisotropic Bianchi type V universe respectively. Therefore, as usual we assume that
a scalar field is the source of dark energy. The energy density and pressure of the scalar field are given by
\begin{equation}
\label{eq37} \rho_{\phi}=\frac{1}{2}\epsilon
\dot{\phi}^{2}+V(\phi)
\end{equation}
and
\begin{equation}
\label{eq38}p_{\phi}=\frac{1}{2}\epsilon \dot{\phi}^{2}-V(\phi),
\end{equation}
where $\epsilon=\pm 1$. $\epsilon=1$ is referred to as
quintessence whereas $\epsilon=-1$ is referred to as phantom. Since the EoS parameter of scalar field is given by
\begin{equation}
\label{eq39}\omega_{\phi}=\frac{\epsilon \dot{\phi}^{2}-2V(\phi)}{\epsilon
\dot{\phi}^{2}+2V(\phi)},
\end{equation}
one can easily find the scalar field and its potential as
\begin{equation}
\label{eq40} \dot{\phi}^{2}=\epsilon(1+\omega_{\phi})\rho_{\phi},
\end{equation}
and
\begin{equation}
\label{eq41} V(\phi)=\frac{1}{2}(1-\omega_{\phi})\rho_{\phi}.
\end{equation}
Now by putting $\omega_{\phi}=\omega^{de}$ and $\omega_{\phi}=\omega^{de}_{eff}$ in eqs. (\ref{eq40}) and (\ref{eq41}), the scalar
field $\dot{\phi}$ and the potential $V(\phi)$ for quintessence and phantom DE models are obtained as
\begin{equation}
\label{eq42} \dot{\phi}^{2}=2\epsilon \left[H^{2}(1+q)-L^{2}a^{-6}-\alpha^{2}a^{-2}-\frac{3}{2}H^{2}\Omega^{m}_{0}a^{-3}-
\frac{1}{2}\Gamma\zeta_{0}\frac{H^{1-2\eta}}{(\Omega^{de})^{\eta-1}}\right],
\end{equation}
and
\begin{equation}
\label{eq43} V(\phi)=2\left[H^{2}(1-q)-\alpha^{2}a^{-2}-\frac{3}{2}H^{2}\Omega^{m}_{0}a^{-3}+\frac{1}{2}\Gamma\zeta_{0}\frac{H^{1-2\eta}}{(\Omega^{de})^{\eta-1}}
\right],
\end{equation}
respectively, where $\zeta_{0}=3^{2-\eta}\xi_{0}$. Note that in above equations for non-viscous (quintessence) dark energy, $\Gamma=0$ whereas
for viscous (phantom) dark energy model, $\Gamma=1$ . As an example, we take $\eta=0.5$ which leads to a power-law expansion for the scale
factor \cite{ref27}. In this case the late time behaviors of $\phi$ and $V(\phi)$, approximately, are given by
\begin{equation}
\label{eq44}\phi \sim \begin{cases}
\lambda &\mbox{ for quintessence} \\
\sqrt{\zeta_{0}}t+\lambda &\mbox{ for phantom}
\end{cases},
\end{equation}
and
\begin{equation}
\label{eq45} V(\phi) \sim \begin{cases}
0 &\mbox{ for quintessence} \\
\zeta_{0} &\mbox{ for phantom}
\end{cases},
\end{equation}
respectively where $\lambda$ is an integration constant. Eqs.(\ref{eq44}) and (\ref{eq45}) clearly show that the scalar field and the
potential of quintessence decrease more faster than those of phantom model. But ultimately when $a\to \infty, \zeta_{0}\to 0$, for
both quintessence and phantom scenario $\phi=\lambda$ and the potential asymptotically tends to vanish.
%%%%%%%%%%%%%%%%%%%%%%%%%%%%%%%%%%%%%%%%%%%%%%%%%%%%%%%%%%%%%%%%%%%%%%%%%%%%%%%%%%%%%%%%%%%%%%%%%%%%%%%%
%%%%%%%%%%%%%%%%%%%%%%%%%%%%%%%% SECTION 3  %%%%%%%%%%%%%%%%%%%%%%%%%%%%%%%%%%%%%%%%%%%%%%%%%%%%%%%%%%%
\section{Thermodynamical Picture of The Dark Energy Model}
The continuity equations for the viscous dark energy could be written as
\begin{equation}
\label{eq46} \dot{\rho}^{de}+3H(\rho^{de}+p^{de})=9\xi H^{2}.
\end{equation}
In a co-moving volume $V$, the total energy density is $U^{de}=\rho^{de} V$. Using this equation in eq. (\ref{eq52}) we
obtain the equation for production of entropy $S^{de}$ in a co-moving volume due to dissipative effects
in a fluid with temperature $T$ as
\begin{equation}
\label{eq47} T\dot{S}^{de}=\dot{U}^{de}+p^{de}\dot{V}=9\xi VH^{2},
\end{equation}
or
\begin{equation}
\label{eq48} \dot{S}^{de}=\left(\frac{T}{V}\right)9\xi H^{2}.
\end{equation}
In case when the density and pressure of the cosmic fluid are functions of the temperature only i.e. $\rho=\rho(T)$ and $p=p(T)$, the
first law of thermodynamics is given by
\begin{equation}
\label{eq49} TdS^{de}=dU^{de}+p^{de}dV= d[(\rho^{de}+p^{de})V]-Vdp^{de},
\end{equation}
or
\begin{equation}
\label{eq50} S^{de}=\frac{V}{T}(\rho^{de}+p^{de})=\frac{V}{T}\left(1+\omega^{de}_{eff}\right)\rho^{de}.
\end{equation}
Since $V=a^{3}$ and the temperature of the event horizon is $T\propto \frac{1}{a}$, using eq. (\ref{eq34}), we can find the equation
of dark energy entropy density as
 \begin{equation}
\label{eq51} S^{de}=c_{\gamma}\left(\frac{\Omega^{de}}{\Omega^{\gamma}}\right)\left(1+\omega^{de}_{eff}\right),
\end{equation}
where $\Omega^{\gamma}=\frac{\rho^{\gamma}}{3H^{2}}$, $\rho^{\gamma}\propto a^{-1}$ and $c_{\gamma}$ is a constant. Inserting
eq. (\ref{eq51}) in eq. (\ref{eq48}) we find
\begin{equation}
\label{eq52} \left(\frac{\dot{S}}{S}\right)^{de}=\frac{3\xi}{(1+\omega^{de}_{eff})\Omega^{de}}.
\end{equation}
Now let us consider two limiting cases at $z\to -1$, i.e $\omega^{de}_{eff}\to -1^{\pm}$. when $\omega^{de}_{eff}\to -1^{+}$, it
means that the dark energy EoS parameter goes to cosmological constant from quintessence region and $\omega^{de}_{eff}\to -1^{-}$, means
that the EoS parameter goes to the cosmological constant from phantom region. If we assume that $(1+\omega^{de}_{eff})$ tends to zero
faster than $\xi$, then we get
Integrating eq. (\ref{eq52}) we obtain
\begin{equation}
\label{eq53} \lim_{\omega^{de}_{eff}\to -1^{+}}\left(\frac{\dot{S}}{S}\right)^{de}\sim + \infty ~~~~~~~~ \mbox{or}~~~~~~~~ S^{de}\to \infty,
\end{equation}
and
\begin{equation}
\label{eq54} \lim_{\omega^{de}_{eff}\to -1^{-}}\left(\frac{\dot{S}}{S}\right)^{de}\sim - \infty ~~~~~~~~ \mbox{or}~~~~~~~~ S^{de}\to 0.
\end{equation}
According to the thermodynamics second law and based on eqs. (\ref{eq53}) and (\ref{eq54}), we conclude that ultimately the dark energy
EoS parameter tends to the cosmological constant from quintessence not phantom. Therefore, although presence of viscosity in cosmic fluid
causes EoS parameter to cross PDL but there should be a transition from phantom to quintessence at late time. It is worth mentioning that
this result is completely independent on whether there is an interaction between dark components or not. Also since in vacuum dominated era
the density and pressure are not functions of temperature, the above equations do not have any application in this case.
%%%%%%%%%%%%%%%%%%%%%%%%%%%%%%%%%%%%%%%%%%%%%%%%%%%%%%%%%%%%%%%%%%%%%%%%%%%%%%%%%%%%%%%%%%%%%%%%%%%%%%%%%%%%%%%
%%%%%%%%%%%%%%%%%%%%%%%%%%%%%%%  SECTION 4  %%%%%%%%%%%%%%%%%%%%%%%%%%
\section{Observational constraints}
From eqs. (\ref{eq25}) and (\ref{eq28}) one can easily find the Hubble parameter $H(z)$ as:
\begin{equation}
\label{eq55} H(z)^{2}=H_{0}^{2}\left[\Omega^{m}_{0}(1+z)^{3}+\Omega_{0}^{r}(1+z)^{4}+\alpha(1+z)^{2}+\beta(1+z)^{6}+\Omega^{de}_{0}(1+z)^{3(1+\omega^{de})}\right],
\end{equation}
where $\omega^{de}$ is assumed to be a constant parameter and $\beta=\frac{L^{2}}{3}$. Since according to \cite{ref28} a constant 
value of $\xi_{0}$
push the EoS parameter to cross phantom divided line (PLD), for the case when cosmic fluid is viscose, the background expansion for
the model reads as
\begin{equation}
\label{eq56} E(z)=\left[\Omega^{m}_{0}(1+z)^{3}+\Omega_{0}^{r}(1+z)^{4}+\alpha(1+z)^{2}+\beta(1+z)^{6}+\Omega^{de}_{0}(1+z)^{3(1+\omega^{de}-\delta)}\right]^{\frac{1}{2}}.
\end{equation}
Here the effective dark energy EoS parameter is $\omega^{de}_{eff}=\omega^{de}-\delta$ (here $\xi_{0}$ is replaced by $\delta$).\\

In what follows, to place observational constraints on parameters space ${\bf P}=\{\Omega^{m},\Omega^{de},\alpha,\beta, \delta, \omega^{de}_{eff},H_{0}\}$,
of our model we use the compilation of 28 Hubble parameter measurements in the redshift range $0.07<z<2.3$ as depicted in Table 1.
The $28 H(z)$ data points have been compiled by Farooq and Ratra \cite{ref29} to find constraints on parameters of some dark energy models
\footnote{As this work completed, we noticed that new 38 Hubble parameter measurements in the redshift range $0.07<z<2.36$ 
have been compiled by Farooq et al \cite{ref30}. Using the $38 H(z)$ data points could make a small change of order of few percent
in our derived parameters.}. Here following their methodology we constrain our model parameter space {\bf P} by
minimizing the following chi-square
\begin{equation}
\label{eq57} \chi^{2}_{H}({\bf P})=\sum_{i=1}^{28}\frac{[H^{th}(z_{i},{\bf P})-H^{obs}(z_{i})]^{2}}{\sigma_{H,i}^{2}},
\end{equation}
%%%%%%%%%%%%%%%%%%%%%%%%%%%%%%%%%%%%%%%%%%%%%%%%%%%%%% Figure 1 %%%%%%%%%%%%%%%%%%%%%%%%%%%
%%%%%%%%%%%%%%%%%%%%%%%%%%%%%%%%%%%%%%%%%%%%%%%%%%%%%%%%%%%%%%%%%%%%%%%%%%%%%%%%%%%%%%%%%%
\begin{figure}[htb]
\centering
\includegraphics[width=15cm,height=15cm,angle=0]{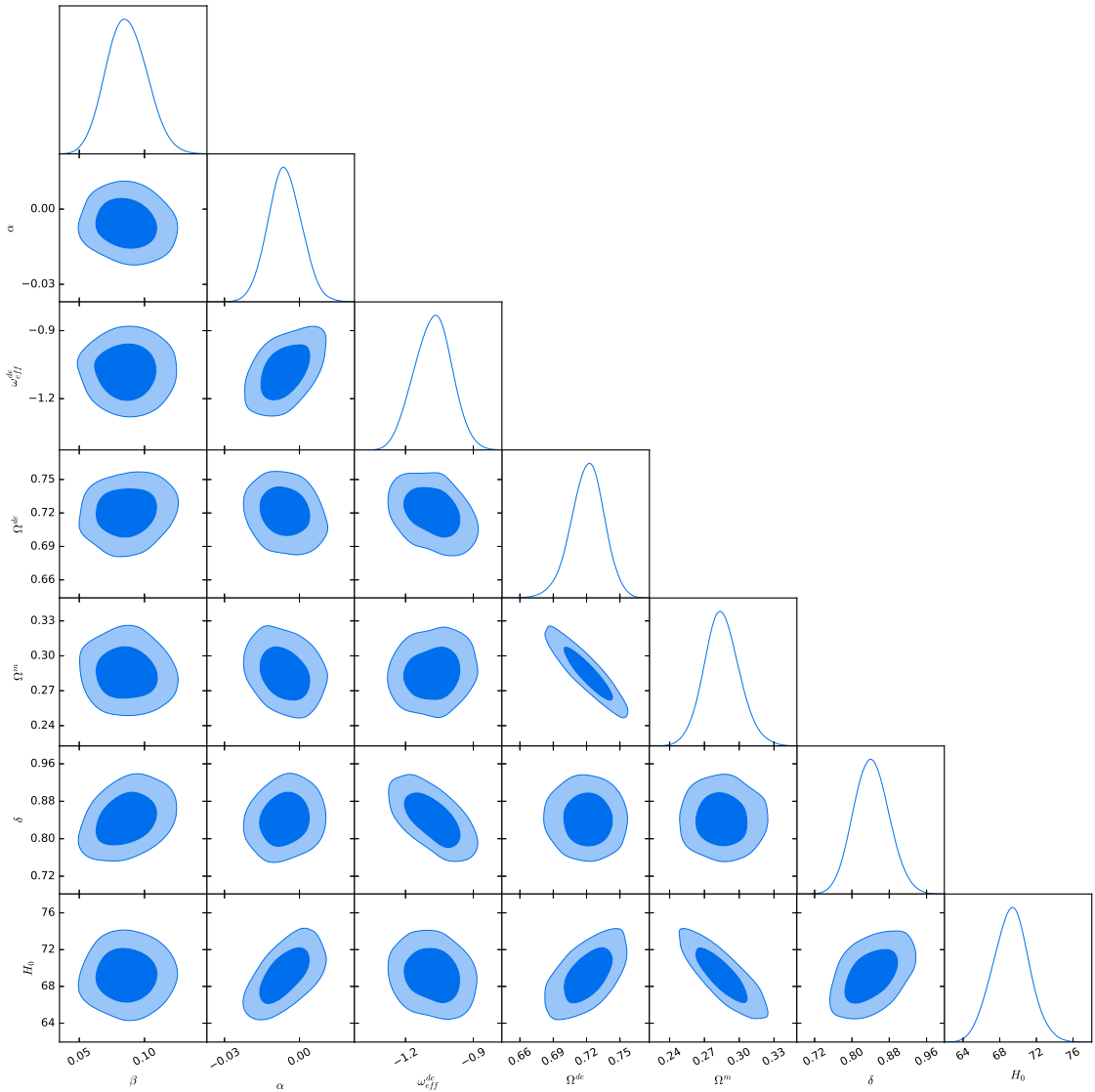}
\caption{One-dimensional marginalized distribution, and two-dimensional contours with $68\%$ CL and $95\%$ CL for the model parameters.}
\end{figure}
%%%%%%%%%%%%%%%%%%%%%%%%%%%%%%%%%%%%%%%%%%%%%%%%%%%%%%%%%%%%%%%%%%%%%%%%%%%%%%%%%%%%%%%%%%%
%%%%%%%%%%%%%%%%%%%%%%%%%%%%%%%%%%%%%%%%%%%%%%%%%%%%%%%%%%%%%%%%%%%%%%%%%%%%%%%%%%%%
%%%%%%%%%%%%%%%%%%%%%%%%%%%%%%%%%%%%%%%%%%%%%%%%%%Table 1%%%%%%%%%%%%%%%%%%%%%%%%%%%%%%%%%%%%%%%%%%%%%%%%%%%%%
%%%%%%%%%%%%%%%%%%%%%%%%%%%%%%%%%%%%%%%%%%%%%%%%%%%%%%%%%%%%%%%%%%%%%%%%%%%%%%%%%%%%%%%%%%%%%%%%%%%%%%%%%%%%%%
\begin{table}[ht]
\caption{Hubble parameter versus redshift data.}
\centering
\begin{tabular} {cccc}
\hline
\hline
$H(z)$    &  $\sigma_{H}$   &  $z$  & Reference\\[0.5ex]
  
\hline{\smallskip}
69 &  19.6 & 0.070   & \cite{ref35}\\

69 & 12 & 0.090   & \cite{ref31} \\

68.6 & 26.2    & 0.120  & \cite{ref35}\\

83  & 8  &0.170 & \cite{ref31} \\

75    & 4 & 0.179 & \cite{ref33} \\

75    & 5 & 0.199 & \cite{ref33} \\

72.9   & 29.6     & 0.200 & \cite{ref35} \\
   
77  &   14   & 0.270  & \cite{ref31} \\
  
 88.8 &   36.6   & 0.280 & \cite{ref35} \\ 
  
 76.3 & 5.6     &0.350  &\cite{ref37}  \\ 
  
 83  & 14 &0.352  & \cite{ref33} \\
  
95   &17  &0.400   & \cite{ref31} \\
  
82.6   &7.8  &0.440   & \cite{ref36} \\
   
97   & 62 & 0.480 &\cite{ref32}  \\
  
104   &13  &0.593  & \cite{ref33} \\
  
 87.9  &6.1  &0.600 & \cite{ref36} \\
  
92   &8  &0.680  &\cite{ref34}  \\
  
97.3   &7  &0.730  & \cite{ref36} \\
  
105   &12 &0.781  & \cite{ref33} \\

125 &17 &0.785  & \cite{ref33} \\
  
90 &40 &0.880  & \cite{ref32} \\
  
117 &23 &0.900  & \cite{ref31} \\ 
  
154  &20 &1.037  & \cite{ref33} \\
  
168 &17 &1.300  & \cite{ref31} \\ 
 
177 &18 &1.430  & \cite{ref31} \\
 
140 &14 &1.530  & \cite{ref31} \\
 
202 &40 &1.750  & \cite{ref31} \\
 
224 &8 &2.300  & \cite{ref34} \\  
  
\hline
\hline
\end{tabular}
\label{table:nonlin}
\end{table}
%%%%%%%%%%%%%%%%%%%%%%%%%%%%%%%%%%%%%%%%%%%%%%%%%%%%%%%%%%%%%%%%%%%%%%%%%%%%%%%%%%%%%%%%%%%%%%%%%%%%%%%%%%%%%%%%%
%%%%%%%%%%%%%%%%%%%%%%%%%%%%%%%%%%%%%%%%%%%%%%%%%%%%%%%%%%%%%%%%%%%%%%%%%%%%%%%%%%%%%%%%%%%%%%%%%%%%%%%%%
where ${\bf P}$ denotes the set of our parameters, $H^{th}$ is the theoretical value of $H(z)$ predicted by our DE model and
$\sigma^{2}_{H,i}$ is the error at $z_{i}$. It is worth to mention that we explore the model parameter space {\bf P} of our DE model
by using the Markov Chain Monte Carlo (MCMC) method package CosmoMC \cite{ref37} and the samples have been analyzed by the aid of
Python package, GetDist \cite{ref38}.\\
%%%%%%%%%%%%%%%%%%%%%%%%%%%%%%%%%%%%%%%%%%%%Table 2%%%%%%%%%%%%%%%%%%%%%%%%%%%%%%%%%%%%%%%%%%%%%%%%%%%%%%%%%%%
%%%%%%%%%%%%%%%%%%%%%%%%%%%%%%%%%%%%%%%%%%%%%%%%%%%%%%%%%%%%%%%%%%%%%%%%%%%%%%%%%%%%%%%%%%%%%%%%%%%%%%
\begin{table}[ht]
\caption{The best fit parameters with 1$\sigma$ and $2\sigma$ confidence level.}
\centering
\begin{tabular} {cccc}
\hline
 Parameter     &   68\% CL  &  95\% CL & Best-Fit Value\\[0.5ex]
\hline
\hline{\smallskip}
$\beta$               & $0.086\pm 0.015$      & $0.086^{+0.032}_{-0.030}$     &  0.085\\

$\alpha$              & $-0.0058\pm 0.0067$   & $-0.006^{+0.013}_{-0.013}$    &  -0.0057\\

$\Omega^{de}$         & $0.721\pm 0.016$      & $0.721^{+0.029}_{-0.032}$     &   0.726\\

$\Omega^{m}$          & $0.285\pm 0.016$      & $0.285^{+0.032}_{-0.029}$     &   0.283\\

$\delta$              & $0.842\pm 0.038$      & $0.842^{+0.076}_{-0.073}$     &   0.844\\

$H_{0}$              & $69.2\pm 2.0$      & $69.2^{+4.0}_{-3.9}$     &  69.3\\

$\omega^{de}_{eff}$   & $-1.079\pm 0.082$     & $-1.08^{+0.16}_{-0.16}$       &  -1.184\\[0.5ex]
\hline
\end{tabular}
\label{table:nonlin}
\end{table}
%%%%%%%%%%%%%%%%%%%%%%%%%%%%%%%%%%%%%%%%%%%%Table 3%%%%%%%%%%%%%%%%%%%%%%%%%%%%%%%%%%%%%%%%%%%%%%%%%%%%%%%%%%%
%%%%%%%%%%%%%%%%%%%%%%%%%%%%%%%%%%%%%%%%%%%%%%%%%%%%%%%%%%%%%%%%%%%%%%%%%%%%%%%%%%%%%%%%%%%%%%%%%%%%%%
\begin{table}[ht]
\caption{The value of $H_{0}$ obtained from different researches.}
\centering
\begin{tabular} {cccc}
\hline
Researchers     &   $H_{0}$  &  Reference \\[0.5ex]
\hline
\hline{\smallskip}
Ade et al (Planck 2015)    & $67.8\pm 0.9$ (at $68\%$)      & \cite{ref39}    \\

Chen \& Ratra              & $68\pm 2.8$  (at $68\%$) &  \cite{ref40}   \\

Sievers et al              & $70\pm 2.4$ (at $68\%$)     & \cite{ref41}    \\

Gott et al            & $67\pm 3.5$  (at $68\%$)    & \cite{ref42}     \\

J. Dunkley et al (CMB)     & $69.7\pm 2.5$   (at $68\%$)   & \cite{ref43}    \\

Aubourg et al (BAO)        & $67.3 \pm 1.1 $  (at $68\%$)    & \cite{ref44} \\

V.Lukovic et al            & $66.5\pm 1.8$  (at $68\%$)   & \cite{ref45}      \\

Chen et al            & $68.4^{+2.9}_{-3.3}$  (at $68\%$)   & \cite{ref46}       \\

Riess et al            & $73.24\pm1.74$  (at $68\%$)   & \cite{ref47}     \\ 

Our model              & $69.2\pm 2.0$ (at $68\%$), $69.2^{+4.0}_{-3.9}$ (at $95\%$) & Present work      \\[0.5ex]
\hline
\end{tabular}
\label{table:nonlin}
\end{table}
%%%%%%%%%%%%%%%%%%%%%%%%%%%%%%%%%%%%%%%%%%%%%%%%%%%%%%%%%%%%%%%%%%%%%%%%%%%%%%%%%%%%%%%%%%%%%%%%
%%%%%%%%%%%%%%%%%%%%%%%%%%%%%%%%%%%%%%%%%%%%%%%%%%%%%%%%%%%%%%%%%%%%%%%%%%%%%%%%%%%%%%%%%%%%%%%%%%
Figure 1 depicts the contour plots for parameters of the Model. The results of our statistical
analysis has been shown in table 2 with $\chi^{2}_{min}=8.0$. Our results clearly show the sensitivity of $\omega^{de}_{eff}$
to the specific choice of the bulk viscosity coefficient ($\delta$) which is restricted as $0.80<\delta<0.88$ at $1\sigma$ error
and $0.74<\delta<0.93$ at $2\sigma$ error. In other word, in absence of viscosity our model only varies in quintessence region whereas
for appropriate values of bulk viscosity coefficient the model cross PDL line. It is interesting to note that the parameter $\alpha$, as
could be expected, plays the role like curvature in FRW cosmology. In other hand, we see that parameter $\beta$ is restricted
as $0.062<\beta<0.15$ at $1\sigma$ error and $0.055<\beta<0.122$ at $2\sigma$ error. Therefore, both $\alpha$ and $\beta$ do not
have significant impact on Bianchi type V behavior at late lime (see figure 2 \& Figure 3 for closer view).
Hence, based on our results we can prove this theoretical property which claim that Bianchi type V behaves like flat FRW model at late time.
The robustness of our fit can be viewed by looking at figure 4. A compression of our $H_{0}$ with those obtained by other researchers
could be seen by looking at table 3. Results of this table clearly show that our $H_{0}$ is in high agreement with $H_{0}$ obtained from CMB
\cite{ref43} at $1\sigma (68\%)$ and the difference with other findings is not so considerable. The reduced chi squares $\chi^{2}_{red}$
(at 68\% confident level) for $\omega CDM$ and $\omega BV$ models are $0.81$ (with $\chi^{2}=18.63$) and $0.76$ (with $\chi^{2}=15.96$)
respectively which shows that although both models are acceptable but $\omega CDM$ model is fitted to the data better
than $\omega BV$ model (please also see figure 4). \\
%%%%%%%%%%%%%%%%%%%%%%%%%%%%%%%%%%%%%%%%%%%%%%%%%%%%%%%%%%%%%%%%%%%%%%%%%%%%%%%%%%%%%%%%%%%%
%%%%%%%%%%%%%%%%%%%%%%%%%%%%%%%%%%%%%%%%%%%%%%%%%%%%%% Figure 2 and 3 %%%%%%%%%%%%%%%%%%%%%%%%%%%
\begin{figure}[htb]
\begin{minipage}[b]{0.5\linewidth}
\centering
\includegraphics[width=8cm,height=6cm,angle=0]{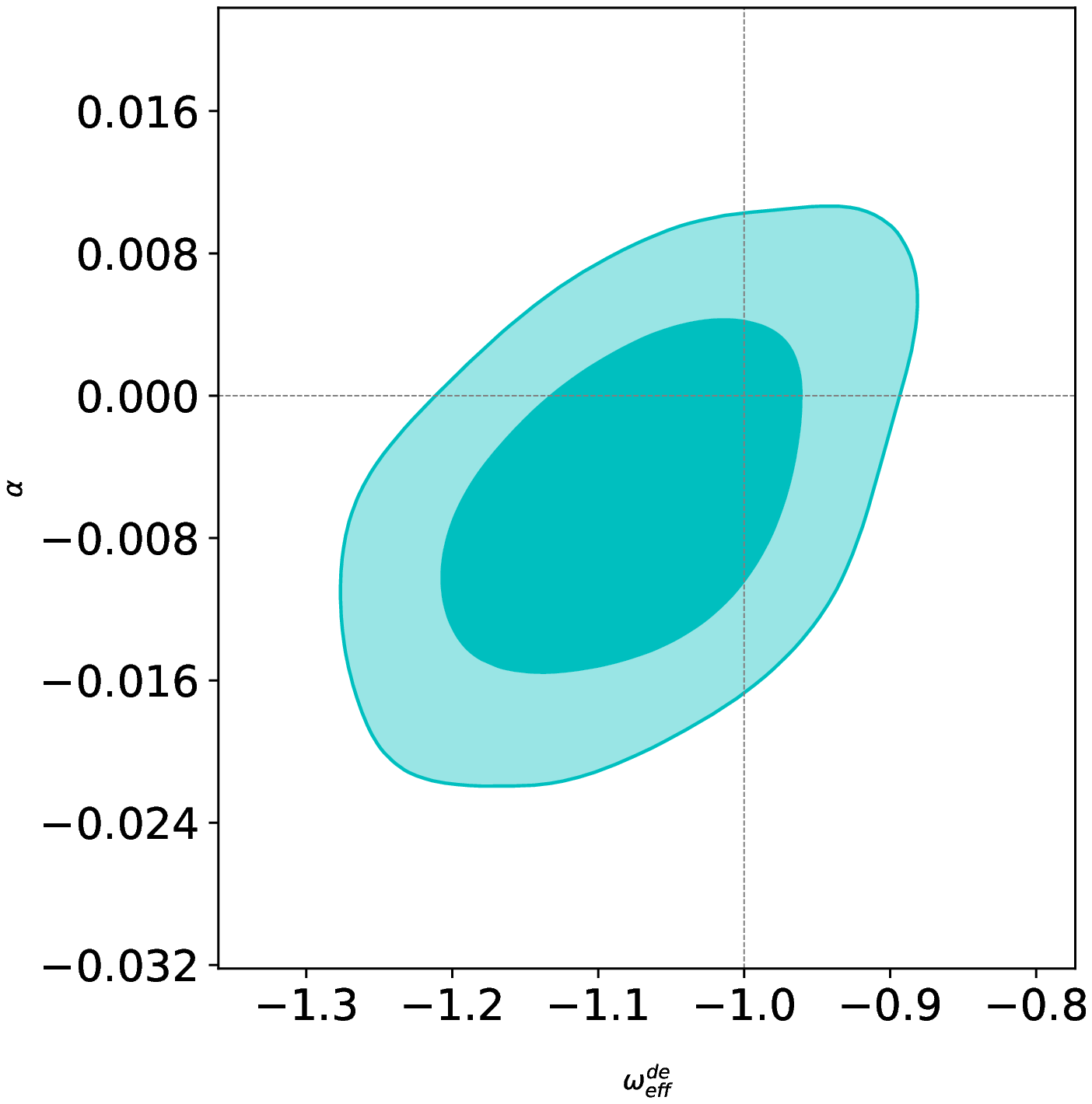} \\
\caption{Two-dimensional contours with $68\%$ CL and $95\%$ CL for $\alpha$ vs $\omega^{de}_{eff}$.}
\end{minipage}
\hspace{0.5cm}
\begin{minipage}[b]{0.5\linewidth}
\centering
\includegraphics[width=8cm,height=6cm,angle=0]{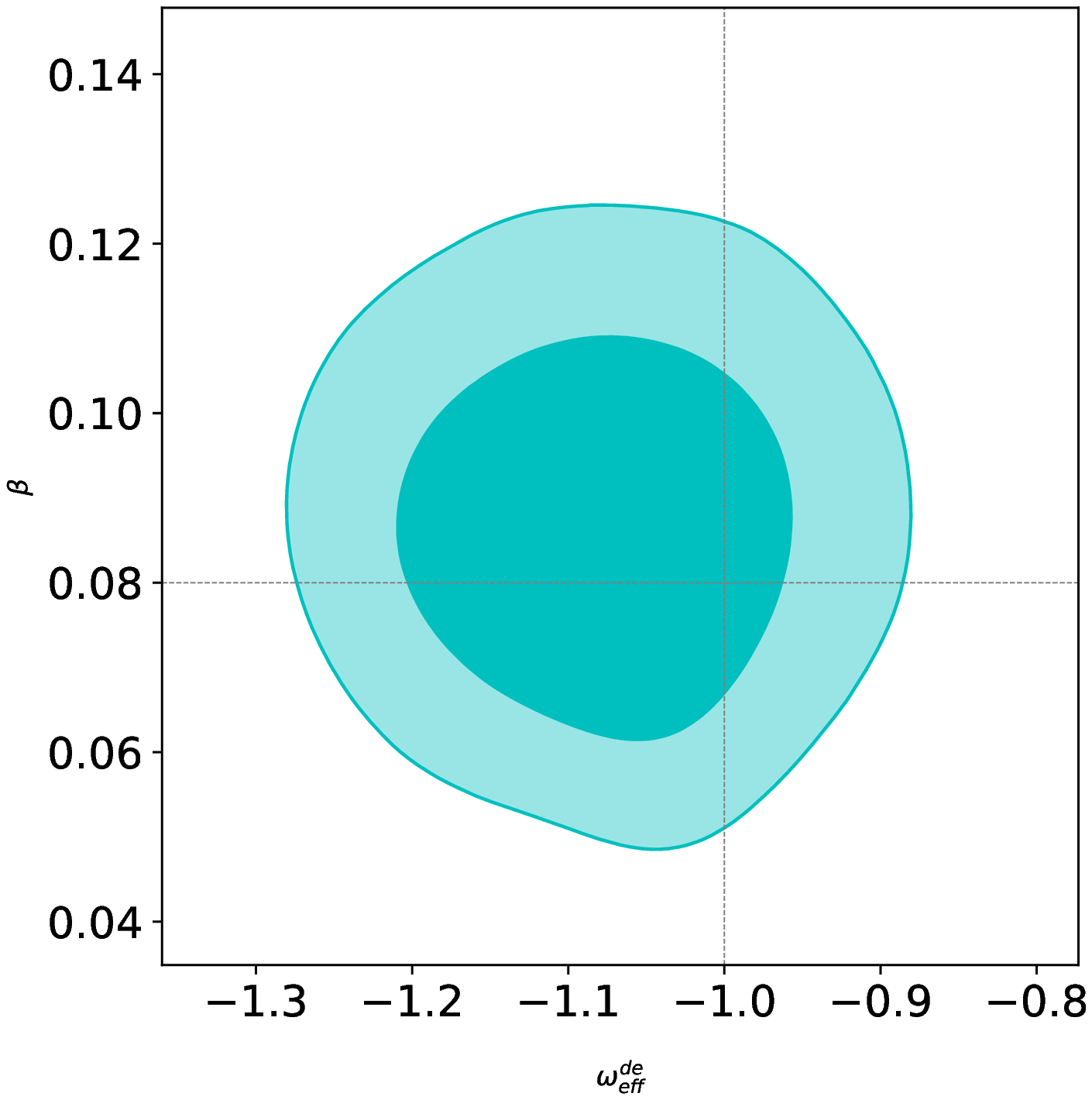}
\caption{Two-dimensional contours with $68\%$ CL and $95\%$ CL for $\beta$ vs $\omega^{de}_{eff}$.}
\end{minipage}
\end{figure}
%%%%%%%%%%%%%%%%%%%%%%%%%%%%%%%%%%%%%%%%%%%%%%%%%%%%%%%%%%%%%%%%%%%%%%%%%%%%%%%%%%%%%%%%%%%

For completeness of our study, here we derive the deceleration-acceleration redshift $z_t$ (this is the redshift at which
the expansion phase changes from decelerating to accelerating). In general, the deceleration parameter is given by
\begin{equation}
\label{eq58} q(z)=-\frac{1}{H^{2}}\left(\frac{\ddot{a}}{a}\right)=\frac{(1+z)}{H(z)}\frac{dH(z)}{dz}-1.
\end{equation}
It is clear that the transition redshift is implicitly defined by the condition $q(z_{t})=\ddot{a}(z_{t})=0$. In case where $L=\alpha=0$
($\omega CDM$ model), from (\ref{eq24}) one can easily find (also see \cite{ref30})
\begin{equation}
\label{eq59} z_{t}= \left(\frac{\Omega^{m}_{0}}{(\Omega^{m}_{0}-1)(1+3\omega^{de})}\right)^{\frac{1}{3\omega^{de}}}-1.
\end{equation}
Also , from (\ref{eq24}), in case where ($\alpha\neq 0, L\neq 0$) we get  
\begin{equation}
\label{eq60} \left(\frac{\ddot{a}}{a}\right)=-2\beta(1+z)^{6}-\frac{1}{6}\left[\Omega^{m}_{0}(1+z)^{3}+(1-\Omega^{m}_{0})(1+3\omega^{de})(1+z)^{3(1+\omega^{de})}\right].
\end{equation}
The transition redshift $z_{t}$ for our $\omega BV$ model could be found by solving the following equation
\begin{equation}
\label{eq61} 2\beta (1+z_{t})^{3}+\frac{1}{6}\left[\Omega^{m}_{0}+(1-\Omega^{m}_{0})(1+\omega^{de})(1+z_{t})^{3\omega^{de}}\right]=0.
\end{equation}
It is interesting to note that in both cases the transition redshift is independent to the parameter $\alpha$ i.e. the curvature of the Model.
Using the best-fit parameters given in table 2 the transition redshift for $\omega CDM$ and $\omega BV$ models are obtained as
$z_{t}=0.652\pm0.105$ and $z_{t}=0.741\pm0.075$ respectively. This result shows that our $\omega BV$ model enters the accelerating phase at earlier
time with respect to the $\omega CDM$ model (see fig 5). Our results are in good agreement with what obtained in refs \cite{ref29,ref30}. Figure 5 indicates the
variation of deceleration parameter, $q$ versus redshift $z$ for both $\omega CDM$ and $\omega BV$ models.
%%%%%%%%%%%%%%%%%%%%%%%%%%%%%%%%%%%%%%%%%%%%%%%%%%%%%%%%%%%%%%%%%%%%%%%%%%%%%%%%%%%%%%%%%%%%%%
%%%%%%%%%%%%%%%%%%%%%%%%%%%%%%%%%%%%%%%%%%%%%%%%%%%%%% Figure 4 and 5 %%%%%%%%%%%%%%%%%%%%%%%%%%%
\begin{figure}[htb]
\begin{minipage}[b]{0.5\linewidth}
\centering
\includegraphics[width=8cm,height=6cm,angle=0]{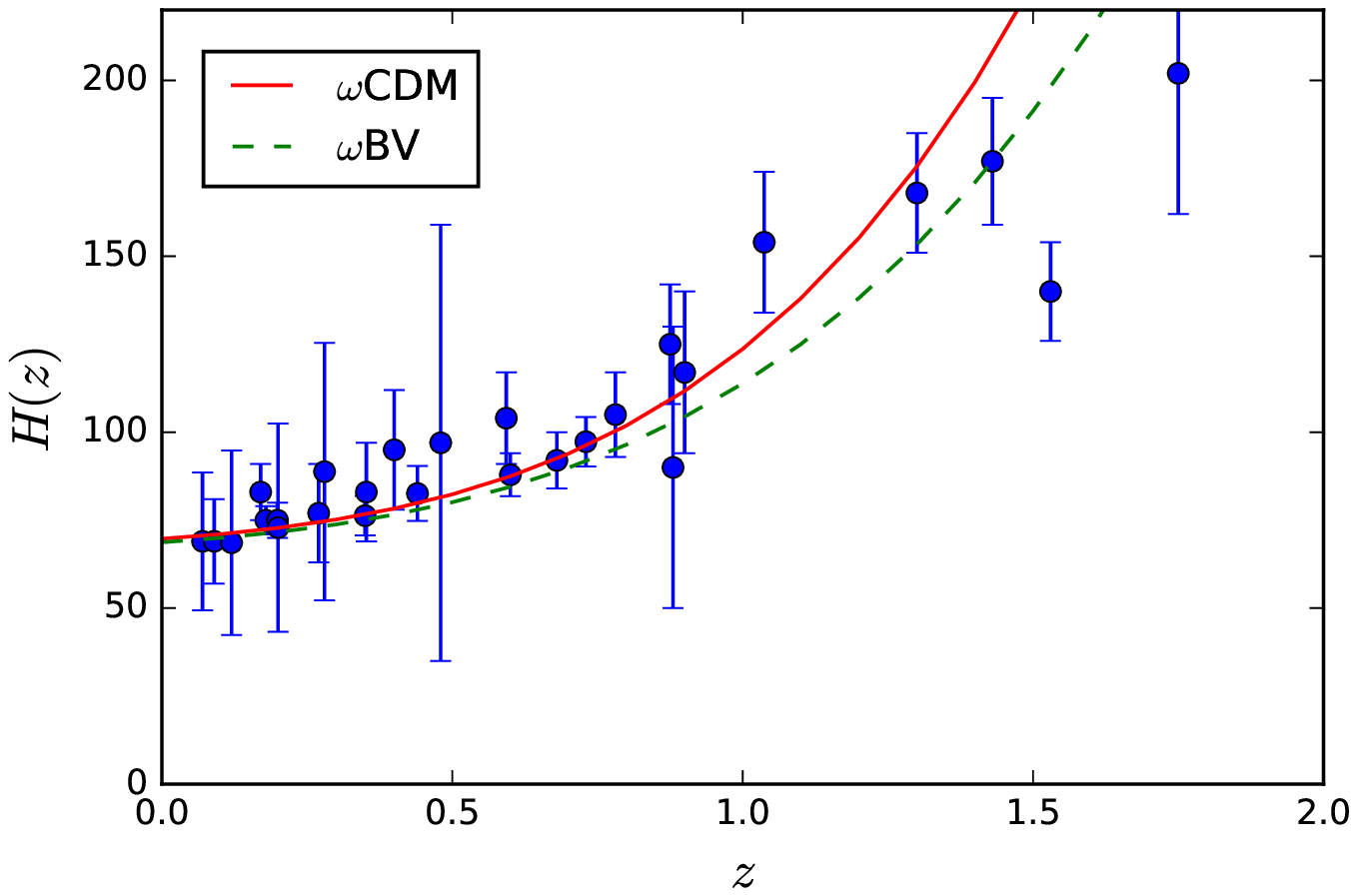}\\
\caption{ The Hubble rate of our model ($\omega CDM$ represent the case when $L=\alpha=0$) versus the redshift $z$.
The points with bars indicate the experimental data summarized in Table 1.}
\end{minipage}
\hspace{0.5cm}
\begin{minipage}[b]{0.5\linewidth}
\centering
\includegraphics[width=8cm,height=6cm,angle=0]{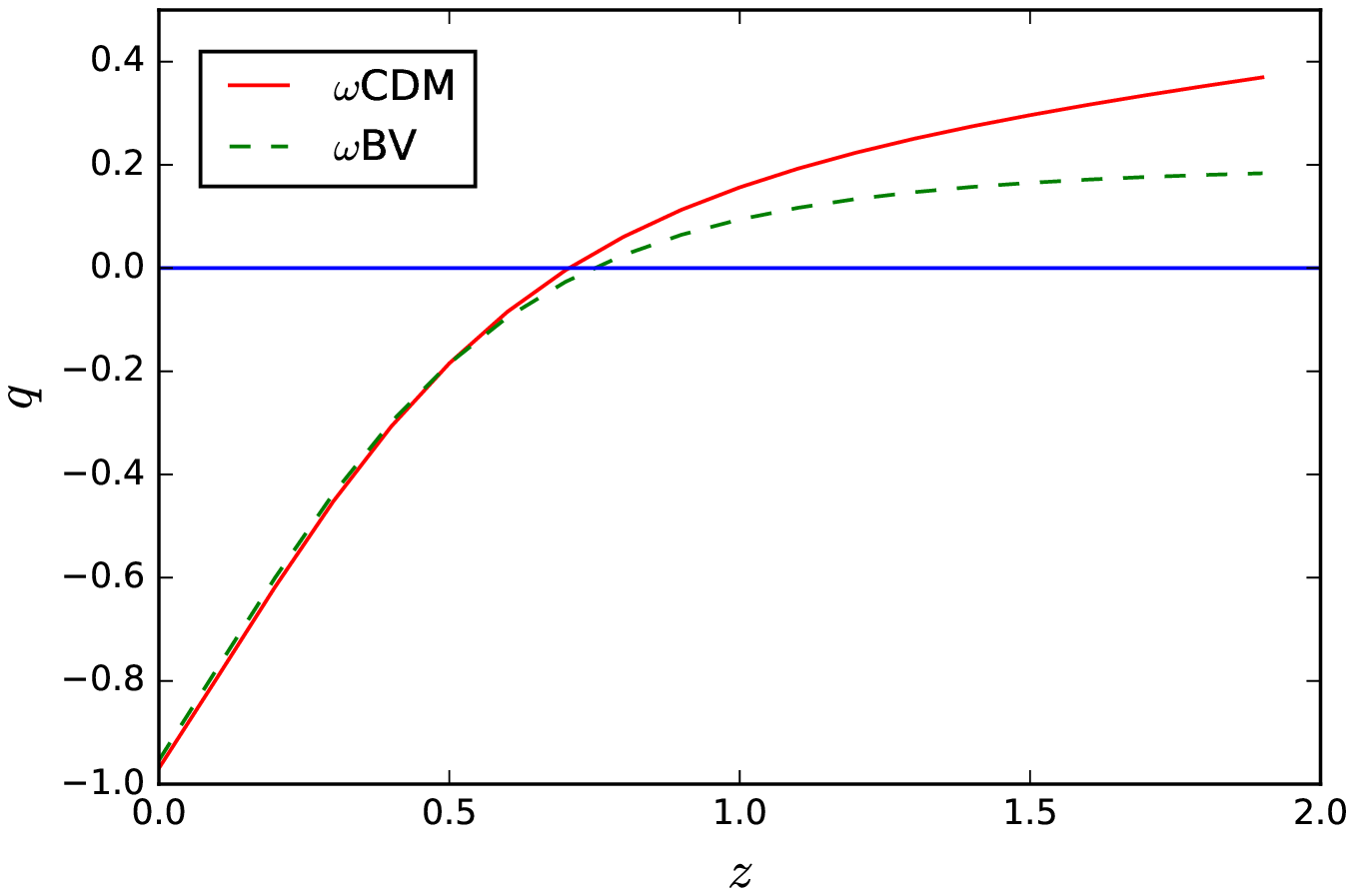}
\caption{variation of deceleration parameter $q$ versus redshift $z$ for both $\omega CDM$ and $\omega BV$ models. The expansion
phase of $\omega BV$ model changes at earlier time with respect to $\omega CDM$ model.}
\end{minipage}
\end{figure}
%%%%%%%%%%%%%%%%%%%%%%%%%%%%%%%%%%%%%%%%%%%%%%%%%%%%%%%%%%%%%%%%%%%%%%%%%%%%%%%%%%%%%%%%%%%%%%%%%%%%%%%%%%
%%%%%%%%%%%%%%%%%%%%%%%%%%%%%%%%%%%%%%%%%%%%%%%%%%%%%%%%%%%%%%%%%%%%%%%%%%%%%%%%%%%%%%%%%%%%%%%%%%%%%%%%%%
\section*{Concluding Remarks}
In this study, we have investigated the behavior of dark energy in the framework of anisotropic Bianchi type V space-time. In general,
we attempted to find the possibility of dark energy EoS parameter to cross PDL line, the correspondence between Dark energy and scalar
fields (phantom as well as quintessence) and the observational constraints on the Model parameters. The main results of this study
can be summarized as bellow.\\
\begin{itemize}
  \item Since recent observations show small departures from isotropy \cite{ref20, ref21} we are motivated to take BV as it could lead to more
  realistic results.
  We derived exact mathematical Friedmann-Like Equations of BV space-time. Based on using some observational data sets, we found that the
  metric parameter $\alpha$ is a negative integer very close to zero which imply that this parameter plays a role like curvature
  in BV cosmology. Also our observational constraint on other metric parameter $\beta$ shows the importance of this parameter and
  its responsibility for inherent anisotropy of BV. Both of these parameter are found to be very small which prove that the FRW metric
  is a special case of BV space-time (i.e. open with $K=-1$). Hence, the study of DE in the scope of anisotropic BV metric is much
  reasonable than FRW metric.
  
  \item We found that although bulk viscosity coefficient $\delta$ is small ( see figure 1) but to cross PDL i.e. transition from
  quintessence to phantom region, the dark sector of the cosmic fluid must be viscous. It is worth noting that, in general, bulk 
  viscosity is a decreasing function of time (see section 4) and hence the viscosity dies out as time is passing and the EoS parameter
  of DE tends to $-1$ as it is necessary condition for any DE model to avoid Big Rip. In section 7 only for simplicity, without loss of
  any generality, we assumed the bulk viscosity to be a constant such as $\delta$.
      
  \item Based on weather cosmic fluid is viscous or not one
can generate phantom or quintessence scenarios in BV space-time. Therefore, as usual we assumed that a scalar field is the source of
dark energy and derived the scalar field $\phi$ and the potential $V(\phi)$ for quintessence and phantom DE Models. It must be noted
that since the energy density of phantom field is unbounded from below, all phantom generally ruled out by ultraviolet quantum
instabilities \cite{ref10} but the viscous DE model proposed here is safe under this problem (see discussion above).
  
  \item We show that, ultimately, the dark energy EoS parameter tends to the cosmological constant from quintessence region not phantom.
  Therefore, although presence of viscosity in cosmic fluid causes EoS parameter to cross PDL but there should be
a transition from phantom to quintessence at late time. It is interesting to note that this result is completely independent
on whether there is an interaction between dark components or not.
\item Our obtained $H_{0}$ is in high agreement with it's value obtained from CMB at $68\%$ confident level. The difference
from other values, as can be seen from table 3, is not so considerable. It is worth to mention that applying new $38 H(z)$ data points compiled in
ref \cite{ref30} can make a small change in our results by a few percent.
\end{itemize}
%%%%%%%%%%%%%%%%%%%%%%%%%%%%%%%%%%%%%%%%%%%%%%%%%%%%%%%%%%%%%%%%%%%%%%%%%%%%%%%%%%%%%%%%%%%%%%%%%%%%%%%%%%
\section*{Acknowledgments}
This work has been supported by the research fund from the Mahshahr Branch, Islamic Azad University under the
project entitled \textquotedblleft Study of The Interaction Between Dark Energy and Dark Matter in Bianchi Type V Universe\textquotedblright.
The author is also grateful to the anonymous referee for valuable comments and suggestions which significantly improved the quality of the
manuscript.

%%%%%%%%%%%%%%%%%%%%%%%%%%%%%%%%%%%%%%%%%%%%%%%%%%%%%%%%%%%%%%%%%%%%%%%%%%%%%%%%%%%%%%%%%%%%%%%%%%%%%%%%%%%%%

\end{document}